\documentstyle[12pt]{article}
\def\ideq{\equiv}
\def\bar{\overline}
\def\grad0{{\stackrel{\circ}{\nabla}}}
\def\d{\delta}
\def\half{\frac{1}{2}}
\def\Lie{{\cal L}}

\begin{document}
\baselineskip=16pt plus 0.2pt minus 0.2pt
\lineskip=16pt plus 0.2pt minus 0.2pt

\begin{center}
 \Large
Energy extremality in the presence of a black hole\\

\vspace*{0.35in}

\large

Rafael D. Sorkin $^{1,2}$ and
Madhavan Varadarajan $^{1,3}$
\vspace*{0.25in}

\normalsize

$^{1}$Department of Physics, Syracuse University, \\
Syracuse, New York 13244-1130, U.S.A.
\\
$^{2}$Instituto de Ciencias Nucleares, UNAM\\
A. Postal 70-543, M\'exico D.F. 04510, M\'exico
 \\
$^{3}$Department of Physics, University of Utah, \\
Salt Lake City, Utah 84112, U.S.A.

\vspace{.5in}
\vspace{.5in}
ABSTRACT

\end{center}

We derive the so-called first law of black hole mechanics for variations
about stationary black hole solutions to the Einstein--Maxwell equations in
the absence of sources.  That is, we prove that
$\delta M=\kappa\delta A+\omega\delta J+VdQ$ where the black hole
parameters $M, \kappa, A, \omega, J, V$ and $Q$ denote mass, surface
gravity, horizon area, angular velocity of the horizon, angular momentum,
electric potential of the horizon and charge respectively.  The unvaried
fields are those of a stationary, charged, rotating black hole and the
variation is to an arbitrary `nearby' black hole which is not necessarily
stationary.  Our approach is 4-dimensional in spirit and uses techniques
involving Action variations and Noether operators.  We show that the above
formula holds on any asymptotically flat spatial 3-slice which extends from
an arbitrary cross-section of the (future) horizon to spatial infinity.
(Thus, the existence of a bifurcation surface is irrelevant to our
demonstration.  On the other hand, the derivation assumes without proof
that the horizon possesses at least one of the following two (related)
properties: ($i$) it cannot be destroyed by arbitrarily small perturbations
of the metric and other fields which may be present, ($ii$) the expansion
of the null geodesic generators of the perturbed horizon goes to zero in
the distant future.)

\pagebreak

\setcounter{page}{1}

\section{Introduction}

Working directly with the Action principle which defines a field theory
typically lends clarity and unity to the basic formal relationships of the
theory.  Thus, in the case of symmetries and conservation laws, the first
variation of the Action $S$, leads directly to the conserved ``charges''
themselves \cite{sorkin}, while its second variation leads to conclusions
of which the following is typical: every time-independent solution of the
field equations is an extremum (critical-point) of the total energy $E$
\cite{schutz}

Two directions of generalization of this last theorem suggest themselves as
natural.  On one hand, one can inquire \cite{schutz}  into the
behavior of the {\it second} variation, or ``Hessian'' of $E$
(corresponding to third variation of $S$ itself); on the other hand, one
can allow for the presence of a spacetime boundary, such as will occur
naturally if one narrows one's attention to the region outside a black hole
horizon.  Clearly, both directions of generalization are relevant to the
stability and (more generally) the thermodynamics of black holes; but the
latter kind of generalization seems to be more readily accomplished, as
well as being a necessary prelude to the former kind in the black hole
case.  This second direction is pursued in the present paper, and
leads---via steps which we now sketch in advance---directly to the
so-called first law of black hole mechanics (or thermodynamics).

A central concept in connection with local transformation groups is
that of the Noether operator, an object which acts (in general as a
differential operator) on an infinitesimal group-generator $\xi$ to
give a spacetime current ${\cal J}^a$ corresponding to the variation
of $S$ induced by $\xi$.  (See equations (1) and (3) below.)
When the group is the diffeomorphism group, $\xi$ is a vectorfield
$\xi^a$, and the Noether operator is correspondingly a tensor
(density of weight 1), with the current being formed as ${\cal
T}^a_b \cdot \xi^b$.  Under appropriate conditions of asymptotic
flatness, a Poincar\'e quotient-group of diffeomorphisms serves as a
global symmetry group of the theory, and---for solutions of the
field equations---the Noether operator yields a corresponding family
of conserved charges when applied to the Poincar\'e generators and
integrated over an asymptotically flat spacelike hypersurface or
``slice''.\footnote
{From a physical point of view, it seems best to interpret such a
global charge, say the linear momentum, as the change in $S$ due to an
infinitesimal variation in which the system under consideration is
translated {\it relative to the environment whose presence is
implicit in the use of asymptotically flat boundary conditions}.
Certainly, genuinely asymptotically flat metrics do not occur in
nature, and the translation or rotation of a subsystem cannot in
practice be meaningfully extended beyond some large but finite
radius.  By taking the existence of the environment more explicitly
into account, it ought to be possible to interpret the conserved
charge as the change in the {\it total} $S$, with the various
divergence terms one customarily adds to the Lagrangian serving only
to allow a nominal splitting of $S$, and therefore of $\delta S$,
into separate parts associated with the approximately isolated
subsystem and with the environment.  The same, ``more realistic''
approach is also important for correctly interpreting the divergent
or conditionally convergent integrals which occur in connection with
the angular momentum and boost conservation laws.  (For the elements
of such an approach see \cite{sorkin}).}

The extremality theorem cited above then follows directly from taking an
additional
variation of the defining equation (1) for the Noether operator: the
integral identity that results from the variation yields the theorem when
evaluated on a spacetime region whose fields are chosen to interpolate
between the unvaried and the varied configurations.  In its maximal
generality, this theorem asserts that any field-configuration which both
solves the equations of motion and is invariant under a given infinitesimal
symmetry, is an extremum of the corresponding conserved quantity; and that
in fact any two of the properties, ``solution'', ``symmetric'',
``extremum'' imply the third.\footnote
{Similar relationships can be derived naturally via Hamilton's equations,
but the Lagrangian approach has the advantage of leading more directly to
objects with a spacetime meaning, such as conserved currents and fluxes
(cf.\cite{chandra} for example).  And of course, a spacetime approach is
particularly well suited to dealing with black holes, whose very definition
involves the global causal structure.}

What happens then if we apply the same considerations, not to the entire
spacetime manifold, but just to the region outside an event horizon?

But first, why is it necessary to exclude the interior region at all?  It
is not that the general theorem ceases to apply just because a horizon is
present, but that, for subtle reasons, it does not furnish the kind of
information one might expect in that case.  With the Schwarzschild metric,
for example, there are two possibilities for what is inside the horizon.
There might be the familiar second asymptotic region joined to the
first by a ``throat'', or there might be no such region if the internal
topology  is such that the ``throat'' leads one back out to the same
asymptotic region from which one came
%
%
(the latter case resulting from the former by suitable identifications.)
In the former case the quantity `$E$' which occurs in the theorem turns out
to be the {\it difference} of the energies seen in the two asymptotic
regions, rather than the effective mass of the black hole, while in the
latter case, the meaning of $E$ is physically appropriate, but the solution
is no longer globally stationary technically, because the
non-simple-connectivity of spacetime
obstructs the extension of the timelike Killing
vector to the interior region as a single-valued field.  In neither case
does one learn anything directly about variations of the physical black
hole energy.  However, if we truncate the manifold at the horizon then the
remaining, external portion of the spacetime possesses both a single
asymptotic region and a globally defined Killing vector,
and we can expect that a suitable generalization of our theorem will
furnish the kind of information we desire.

Consider then, an arbitrary variation of the metric in the neighborhood of
an asymptotically flat slice, $\Sigma$
which meets the horizon in an arbitrary two-dimensional cross-section.
(The freedom to choose the cross-section freely is important, because the
ability to push such a section forward in time appears to be
crucial to an understanding of the {\it second} law of black hole
thermodynamics; see the further remarks in the conclusion section below.)
As long as the variation stays
away from the horizon, the general theorem applies exactly as before, and
we conclude that {\it the energy of a time-independent black hole is an
extremum against variations of the fields which vanish in a neighborhood of
the horizon}.  From this alone, it follows immediately that for arbitrary
variations (including non-stationary ones), {\it the change in the total
energy can depend only on the behavior of the variation on the horizon
itself}, or more precisely, on its behavior near the 2-surface where the
horizon meets the slice on which the energy is being evaluated.

We have come this far on the basis of  reasoning valid for any field
theory.  To
discover specifically which horizon-variables determine $\delta E$, we
will need to use the concrete form of the Action $S$.  (One might think
that, in addition to horizon variables, the choice of spacelike hypersurface
might enter
into $\delta E$, but that cannot be true, because the energy of the varied
configuration does not depend on the slice on which it is
evaluated---assuming that the variation itself does not lead out\footnote
{The Noether operator allows us to define a generalized energy, even for
field-configura\-tions which do not satisfy the Einstein equation, and we
will need to use this feature in our derivation of $\delta E$.  Thus, it
may be of some interest that hypersurface independence for this more
general type of variation still follows from the identity on which the
proof is based (i.e. $\delta E$ can depend on the choice of slice only
insofar as this influences the cross-section in which the slice meets the
horizon).}
of the space of solutions.)
For simplicity, consider the special case of a static, uncharged black hole
in pure
gravity, and
let $\xi^a$ be the asymptotically timelike Killing vector of the unvaried
solution.  Then, the integral for $\delta E$ that evaluated to zero in the
absence of a boundary, becomes with a boundary present, the net flux of a
fictitious energy-current (that of the Einstein tensor of the interpolating
metric)
across the horizon.  This flux can be evaluated entirely straightforwardly
using the Raychaudhuri equation, and yields the familiar expression $\kappa
\delta A$, where $A$ is the horizon area and $\kappa$ its surface gravity.
We thus arrive at the first law for non-rotating uncharged black holes.

The more general case of a charged and/or rotating black hole is equally
straightforward, and needs no special elaboration in this introduction,
except to mention that, in dealing with angular momentum, we will need the
analog for rotation of the ``asymptotic patching'' lemma employed in
\cite{schutz} for the case of translation.  This is derived in an appendix
(in a version strengthened significantly over that of \cite{nahmad}, the
improvement being made possible by our imposition of the so-called ``parity
conditions'' on the asymptotic fields in conjunction with the use of an
improved Noether operator over that used in \cite{nahmad}).

The only sightly tricky point in the derivation will be locating the
horizon of the varied metric, a task which in general would demand
knowledge of the entire future of the varied solution.  In fact, what we
will need for our derivation is only that the expansion $\theta$ be of
second order in the perturbation when evaluated on the correctly identified
varied horizon.  Although ``local'' in itself, this assertion refers
implicitly to the entire future, and we do not prove it herein.  Instead,
we reduce it to either of two assumptions asserting the stability (in a
certain sense) of the black hole horizon.  Such assumptions are common in
discussions of black hole dynamics, and may perhaps be viewed as special
cases of ``cosmic censorship''.  Were they to fail, black hole spacetimes
would have to be understood as very special types of mathematical solutions
without relevance to reality.

The layout of the remainder of this paper is as follows. In section 2 we
recall the definition of the Noether operator for gravity and it's relation
to conserved quantities  \cite{sorkin,schutz,chandra}.  The rest
of the main body of the paper is devoted to the proof of the first law of
black hole mechanics for variations from stationary `gravitoelectric'
spacetimes to nearby (not necessarily stationary) gravitoelectric spacetimes
(by `gravitoelectric', we intend gravity coupled to the source-free
electromagnetic field).  In section 3 we sketch the idea for the proof
based on an `extremum' identity and set the stage for the proof with some
technical remarks.  Section 4 gives the proof for ``vacuum'' solutions and
section 5 generalizes it to the gravitoelectric case.  In the concluding
section, we discuss the significance of the result obtained and consider
its relation to earlier work. In the appendices, the asymptotic falloffs of
the
metric and the electromagnetic potential are specified, and a lemma crucial
to our proofs is proved.

Since all variations will be about stationary solutions with vanishing spatial
momentum $\vec{p}$, we will have
 $\delta E=\delta M$ to first order, where $M$ is the
mass of the black hole (compare the relation
$E^{2}=M^{2}+{\vec{p}}^{2}$).  From now on we will denote first order
variations in energy  by $\delta M$, and often call
the energy itself $M$.

For a ``minimalist'' account of what follows, see the first part of
reference \cite{winnipeg}.

\section{The Noether operator for gravity}

For the geometrical invariances of classical field theories, the
`Noetherian' relation between symmetries and conservation laws can be
codified in terms of a differential operator ${\cal T}_{b}^{a}$ which acts
on an arbitrary diffeomorphism-generating vector field $\xi^{a}$ to produce
a conserved current ${\cal T}_{b}^{a}\cdot \xi^{b}$.  One can associate
such a Noether operator to any first order Lagrangian density ${\cal L}[Q]$
via the identity \cite{sorkin,schutz}
\begin{equation}
\delta S \:= \int_{\Omega}(\delta{\cal L}/\delta Q)\delta Q d^{n}x
   \:+\: \int_{\partial{\Omega}}f{\cal T}_{b}^{a}\cdot \xi^{b}d\sigma_{a}
\end{equation}

\noindent In this equation, the $Q$ are the dynamical field variables,
$S=\int_{\Omega}{\cal L} d^{n}x$ is the action, $(\delta{\cal L}/\delta Q)$
is it's variational derivative, and $\delta S$ is the result of a variation
in which (i) the variables $Q$ `change in place' according to
\begin{equation}
\delta Q \:=\: -f{\cal L}_{\xi}Q
\end{equation}
(${\cal L}_{\xi}$ being the Lie-derivative), and (ii) the region of
integration shifts its boundary by an amount $f\xi^a$. Notice that the
arbitrary scalar field $f$ in (1) and (2) helps to reduce the ambiguity of
${\cal T}_{b}^{a}$, which, for $f\ideq 1$, would remain a
solution of (1) after supplementation with an arbitrary total divergence
formed from $\xi$ and the fields $Q$. It is implicit in (1) that
${\cal T}_{b}^{a}={\cal T}_{b}^{a}[Q]$ depends locally on $Q$; and (1)
is  required to hold for arbitrary $\xi$,
$f$ and $\Omega$ {\em  whether or not Q solves the equations of motion}.

The expression  ${\cal T}_{b}^{a}\cdot \xi^{b}$ can be integrated to produce
conserved quantities or `charges' associated with the asymptotic symmetries of
an isolated system in a manner discussed in \cite{sorkin}, wherein the
interpretation of these global charges as variations of the total action is
also explained.
\footnote
{For $\xi$ corresponding asymptotically to a rotation, the integral
defining $\delta S$ converges only conditionally and only to first order in
the variation $\delta$. This obstacle to giving angular momentum a precise
meaning derives from the fact that $S$ itself converges only conditionally
in general. It can be overcome by redefining $S$ as the Action inside a
sufficiently large sphere, and taking limits as the size of the sphere goes
to infinity}
Without going into further detail here, let us remark that the
interpretation is valid when $\xi$ is an exact symmetry of the background
fields, and the presence of the function $f$ in (2)
is what allows the variations in
question to remain compatible with the asymptotic flatness of the metric
$g_{ab}$ at large radii. If $\xi^a$ generates time translation then
$\int d\sigma_{a} {\cal T}_{b}^{a}\cdot \xi^{b}$ will be the total energy;
if it
generates a rotation it will be an angular momentum component, etcetera.

The form of $S_{grav}$ which leads to the Noether operator we use for
gravity is the first order form obtained from the covariant action
$(1/2)\int RdV$ via  integration by parts. To define it in an intrinsic
manner we can introduce a background connection  $\stackrel{\circ}{\nabla}_a$
(say torsion free) and let $\Gamma^{a}_{bc}$ represent the difference
therefrom of the metric's connection $\nabla_{a}$. (In a specific
coordinate system, if
  $\stackrel{\circ}{\nabla}_a$
is chosen as the coordinate derivative operator, then
$\Gamma^{a}_{bc}$
will just be the Christoffel symbol as usually defined.) The first order
Action is then
$S_{grav} \:=\: 1/(2k)\int dV
(\Gamma^{a}_{bc} \Gamma^{b}_{ad}-  \Gamma^{a}_{ab} \Gamma^{b}_{cd}
+ \stackrel{\circ}{R}_{cd})g^{cd}$,
where
$\stackrel{\circ}{R}_{cd}$
is the Ricci tensor of   $\stackrel{\circ}{\nabla}_a$ and $k=8\pi G$ (in 4
spacetime dimensions).
The Noether operator which answers to $S_{grav}$ turns out to be that
acting as follows \cite{sorkin,katz}:
\begin{equation}
  T^{a}_{b} \cdot \xi^{b}\:=
     \:-G^{a}_{b} \xi^b
        +\nabla_{b} (\nabla^{[a}\xi^{b]}+\omega^{[a}\xi^{b]})
\end{equation}
where $T^{a}_{b}=(-g)^{-1/2}{\cal T}^{a}_{b}$, $\omega^{a}=
\Gamma^{a}_{bc}g^{bc} - \Gamma^{b}_{bc}g^{ac}
= (g)^{-1} \grad0_b (- g g^{ab})$,
and
$x^{[a}y^{b]}$ = $1/2[x^{a}y^{b} -x^{b}y^{a}] $.
We also define
$$
   W^{ab}_c\cdot\xi^c=(\nabla^{[a}\xi^{b]}+\omega^{[a}\xi^{b]}) \eqno{(3a)}
$$

\noindent
{\it Note on Notation}: In the various integrals which appear, we shall
incorporate all the density weights into the volume elements which will
accordingly be scalars (of density weight 0).  We write the various volume
elements thusly:
$dV$ denotes the element of 4-volume, $dS_a$ denotes the volume element on
a 3-dimens\-ion\-al submanifold and $dS_{ab}$ that on a 2-dimensional
submanifold (which will always be topologically a 2-sphere).  We also
absorb into the volume elements $dS_a$, etc. the orientations which
implicitly occur in Stokes' Theorem, although in this case the received
notation does not allow us to indicate which orientation is being taken.  In
general relations like equation (9) we will adhere to the rule that the
boundary of a region is to be oriented {\it outward}.  In integrals over
separate spacelike or null hypersurfaces like the integrals over $\Sigma$
in (13) and (14), we will orient the surfaces upward (as if they were
regarded as the boundaries of their pasts).  When the surface is a portion
$H$ of a black hole horizon, this has the consequence that $H$ is oriented
``inward'' (as if regarded as the boundary of the region of spacetime
outside the black hole).  Other orientations will not be made explicit, but
can be deduced from the context.

With these conventions, the energy $E$ of a gravitational field evaluated
on a spatial asymptotically flat slice $\Sigma$ without boundary is given
by
\begin{center}
       $ -E = (1/k) \int _{\Sigma} T^{a}_{b}\cdot \xi^{b} dS_{a}$
\end{center}
\noindent
where $\xi$ is an asymptotic time translation and $T^{a}_{b}$ is an
ordinary tensor operator, not an object of density weight 1. Note that when
the vacuum Einstein equations are satisfied, we have from (3) and (3a),
\begin{equation}
      E = -(1/2)\int_{\partial \Sigma} W^{ab}_c\cdot\xi^c dS_{ab}
\end{equation}
where $\partial \Sigma$ is the ``2-sphere at spatial infinity" and
$k=8\pi G$ is set equal to $1$.

The assumed fall off conditions on the metric are given in the Appendix.

\section{Preparation for the proof}
For a more detailed exposition of the ideas in this section see
\cite{schutz,ontimunu}.

\subsection{The ``extremum identity'' for variations about a stationary
solution, and its application to black hole spacetimes}

We notice   that for a variation which corresponds to a
pure diffeomorphism
generated by $\xi^{a}$ (which corresponds to putting  $f=1$
in equation(1)) we have for the gravitational action in (3)
\begin{equation}
\delta S_{grav}:= \: \delta_{\xi} S_{grav}
=\int_{\Omega}  dV (1/2) G^{ab} {\cal L}_{\xi}g_{ab} +
\int _{\partial \Omega} T^{a}_{b} \cdot \xi^{b} dS_{a}
\end{equation}

\noindent But $S_{grav}$ is invariant under diffeomorphisms which preserve
the background connection
$\stackrel {\circ}{\nabla}$.
Let us assume that $\xi$ preserves the background connection; then
\begin{equation}
\delta_{\xi}S  =  0
\end{equation}
\begin{equation}
\Rightarrow
\int_{\Omega} dV (1/2) G^{ab} {\cal L}_{\xi}g_{ab} +
\int _{\partial \Omega} T^{a}_{b}\cdot \xi^{b} dS_{a}  = 0
\end{equation}

\noindent Note that this identity holds for arbitrary field
configurations,
which do not need to be solutions to the field equations. We now take the
variation of (7) under an arbitrary variation $\delta g_{ab}$
of the metric. We get
\begin{equation}
 \int_{\Omega}(1/2)\delta ( dV G^{ab}) {\cal L}_{\xi}g_{ab} +
 \int_{\Omega} dV (1/2) G^{ab} \delta( {\cal L}_{\xi}g_{ab}) +
 \delta\int _{\partial \Omega} T^{a}_{b}\cdot\xi^{b} dS_{a} \;\;=\;\;0
\end{equation}

Let us further assume that we are varying about a metric which is both a
solution
to the field equations ($G_{ab}=0$) and stationary with respect to
$\xi^a$ (${\cal L}_\xi g_{ab}=0$).  Then the above equation reduces to
\begin{equation}
\delta\int _{\partial \Omega} T^{a}_{b} \cdot \xi^{b} dS_{a} \;=\;0
\end{equation}
which we shall call the {\it extremum identity}.

 For the proof, we will take $g_{ab}$ to be a stationary black hole
solution to the vacuum field equations with killing vector field $\xi^{a}$,
and we will choose $\Omega$ as the region bounded between the horizon
`$\partial B$' of the stationary black hole and two asymptotically flat
spatial slices, $\Sigma_{i}$ and $\Sigma_{f}$ ($\Sigma_f$ being to the
future of $\Sigma_i$), each of which intersect the horizon at one ``end"
and go out to spatial infinity at the other ``end".  Also, we will choose
$\xi^{a}$ to be that Killing field of the stationary  solution
which is null on the horizon.  (The general Kerr-Newman metric always
admits such a Killing vector; the horizon ``rotates rigidly''.)  Then
\begin{equation}
\xi^{a}\;=\;t^{a}\;+\;\omega\phi^{a}
\end{equation}
where $t^a$ and $\phi^a$ are the Killing vectors of the stationary solution
which are asymptotically generators of time translation and rotation
respectively; and $\omega$ is the angular velocity of the stationary horizon
with respect to infinity.  We aim to reproduce the first law
with $E$, $A$, etc. evaluated on the slice $\Sigma_{f}$.

To that end, a suitable choice of ``gauge'' for $\d g$ will be very
helpful.  The gauge-freedom arises in the following, well-known manner.  If
$g$ is a metric on a manifold $N$ and $\bar g$ is a metric on $\bar N$,
then the pairs $(N,g)$ and $(\bar{N},\bar{g})$ may be said to be
``infinitesimally close'' iff there exists a diffeomorphism
$\Phi:N\rightarrow \bar{N}$, such that $\Phi^{*} \bar{g} =g+\delta g$ where
$\delta g$ is infinitesimal.\footnote
{The notion of infinitesimal can be made precise in the usual way: replace
 $\Phi$ and $\bar g$ by smooth parameterized families $\Phi(\lambda)$ and
 ${\bar g}(\lambda)$ with $\Phi(0)^* {\bar g}(0) = g$, and define
 $\d g :=
  \partial[\Phi(\lambda)^*{\bar g}(\lambda)]/\partial\lambda |_{\lambda=0}
  d\lambda$.  Similarly, we interpret (b) below (for example) to mean that
the horizons coincide for the members of some 1-parameter family of
solutions whose derivative is $\delta g_{ab}$}
Such a diffeomorphism $\Phi$ (call it `allowable') is not unique: if $\Pi$
is allowable and $\Psi$ is any diffeomorphism infinitesimally close to the
identity, $\Pi \circ \Psi$ is also allowable.  We will use this freedom in
the specification of $\Phi$ in (b), (c) and (d) below.

We now divide the 4-manifold outside $\partial B$ into 3 parts:\\
\quad{$\bullet$} $\Sigma_i$ and its past\\
\quad{$\bullet$} $\Sigma_f$ and its future\\
\quad{$\bullet$} The region $\Omega$ between $\Sigma_i$ and $\Sigma_f$\\
and we contemplate a preliminary choice of $\delta g_{ab}$ as follows.

\noindent (a) On $\Sigma_i$ and to it's past, $\delta g_{ab}=0$

\noindent (b) On $\Sigma_f$ and to it's future
${(g_{ab}+ \delta g_{ab})}$
is a black hole solution to the linearized field equations with  mass
$M+\delta M$
and angular momentum $J+\delta J$.
We assume,
without loss of generality, that the
horizons of the varied  solution and of the stationary unvaried
solution coincide in this region.

\noindent
(c) In $\Omega$: we require that the part of $\partial B$ between
$\Sigma_i$ and $\Sigma_f$ be a null surface for the metric ${(g_{ab}+
\delta g_{ab})}$.  We shall henceforth refer to this portion of the horizon
  of $g+\d g$ as $H$.

\noindent
(d) On $\partial B$ it is possible to ``line up'' the null
geodesic generators  $k^{a}$, for $g_{ab}$ with those for
$g_{ab}+\delta g_{ab}$, so that $\delta k^{a}$ is proportional to $k^a$.
This alone does not imply that the affine parameters of the
unvaried and varied null geodesic fields agree on $H$. In general they
will differ by  infinitesimal amounts. But it is always possible to use
the infinitesimal diffeomorphism freedom
whose existence we emphasized above,
to identify slices in the unvaried spacetime with slices in the varied
spacetime, in such a way that the affine parameters actually do agree. Thus,
on $H$, we can always arrange for $\delta k^{a}=0$, without in any way
constraining the location of $\Sigma_f$ with respect to the varied
metric.

Based on our choice of $\d g$, we shall end up proving the following
statement (for the vacuum case): Given a stationary vacuum black hole
spacetime $N$ with metric $g$ and an infinitesimally nearby pair
$(\bar N, \bar g)$, there exists an allowable diffeomorphism $\Phi_0$,
which maps $\bar g$ to $g + \d g$ and $\bar{\Sigma}$ (a slice in the
exterior of the horizon of the varied spacetime) to an ``infinitesimally
nearby'' slice $\Sigma$ such that the first law holds on $\Sigma$ with
regard to the infinitesimal variation $\delta g$.
But, since {\it all} slices in the stationary spacetime have the same
horizon area, we will have actually proved the first law in the context of
an arbitrary allowable diffeomorphism, not just $\Phi_0$.
Similar considerations apply to the gravito-electric case.

\vspace{4mm}

We choose our background connection $\stackrel{\circ}{\nabla}$, so that it
is compatible near spatial infinity with the flat metric $\eta_{ab}$ which
we will use to define our fall off conditions.  We also choose it
so that it is globally Lie derived by $\xi$.
(For example we could use
${\stackrel{\circ}{\nabla}} =
 (1-\lambda){\stackrel{\circ}{\nabla}}_{flat} +\lambda{\nabla}_{stationary}$
with $\lambda=1$ on $H$, $\lambda=0$ at spatial infinity and
${\cal L}_{\xi}\lambda =0$.)


Note that the requirement that $g+\delta g$ on and to the future
of $\Sigma_f$ be
 a solution with mass $M+\delta M$ and angular momentum $J+\delta J$
different in general from the mass $M$ and angular momentum $J$ of the
stationary solution implies that {\it in the region} $\Omega$,
$g+\delta g$ {\it cannot be a solution to the field equations}.

\subsection{ An important technical point: asymptotic patching}

Although we would like to make a variation satisfying the conditions (a),
(b) (c) and (d) of section III.1 above, we notice that such a variation
would violate the asymptotic fall off conditions on the metric (for these
conditions see the appendix). We can see this in the following way: The
mass-information for a solution to the field equations is in the `$1/r$
part' of the metric. Now, the mass of the initial and final field
configurations on the slices $\Sigma_i$ and $\Sigma_f$ are $M$ and $M +
\delta M$ respectively. So in between these slices near spatial infinity
the $1/r$ part has to be time dependent. But this means that
$\partial_{t}(g+\delta g)$ will fall off as $1/r$ and not $1/r^2$ as
required. Similar considerations apply to `angular momentum information'.
Hence, the fall off conditions in the appendix would have to be violated.
If this happened, the variations of the action $S_{grav}$ would be ill
defined, meaning that we could not use the variation outlined in section
III.1.  This is the technical problem.
To avoid it, we employ the following trick, along
the lines of the trick performed in \cite{schutz}

First recall that our aim is to produce from the extremum identity
the explicit horizon term to which we know that $\delta M$ must reduce.
 Consider, then, an asymptotically flat spatial slice $\Sigma \subset \Omega$
extending from $H$ to spatial infinity.  For an arbitrary black hole
{\it solution} to the field equations, the energy $M$ evaluated on the spatial
slice $\Sigma$ is given by equation (4) as
$M=-(1/2) \int_{S^{2}_{\infty}} W^{ab}_c\cdot t^c dS_{ab}$
(where $S^{2}_{\infty}$ is the 2-sphere at spatial infinity and $t^a$ is
the asymptotic time translation vector field).

 We can extend the definition of the energy to {\em non-solutions}  by
setting
\begin{eqnarray}
 -M & = &  \int_{S^{2}_{\infty}} \frac{1}{2} W^{ab}_c\cdot t^c dS_{ab}
         - \int_{\Sigma} G^{a}_{b} t^{b} dS_a \nonumber\\
 \;\; & = &
 (\int_{S^{2}_{\infty}} - \int_{S^{2}_{\Sigma \cap H}})
          \frac{1}{2} W^{ab}_c\cdot t^c dS_{ab}
 + \int_{S^{2}_{\Sigma \cap H}} \frac{1}{2} W^{ab}_c\cdot t^c dS_{ab}
 - \int_{\Sigma} G^{a}_{b} t^{b}dS_a
\end{eqnarray}
where $S^{2}_{\Sigma \cap H}$ denotes the 2-sphere in which $\Sigma$
meets $H$.
Of course this agrees with the energy  {\it on} a solution because the
Einstein
tensor vanishes on a solution.   Similarly we can  extend
the definition of the angular momentum to non-solutions by setting
\begin{equation}
  J\;=\;
  \int_{S^{2}_{\infty}} \frac{1}{2} W^{ab}_c\cdot\phi^c dS_{ab}
  \;-\; \int_{\Sigma} G^{a}_{b} \phi^{b}dS_a
\end{equation}
(where $\phi^a$ is  the asymptotic rotational vector field).
In view of Stokes theorem and eq. (3), our definitions of energy and
angular momentum extended to non-solutions are, respectively:
\begin{equation}
-M\;=\; \int_{\Sigma} T^{a}_{b}\cdot t^{b}dS_{a}
     \;+\; \int_{S^{2}_{\Sigma \cap H}} \frac{1}{2} W^{ab}_c\cdot t^c dS_{ab}
\end{equation}
and
\begin{equation}
J\;=\; \int_{\Sigma} T^{a}_{b}\cdot\phi^{b}dS_{a}
     \;+\; \int_{S^{2}_{\Sigma \cap H}} \frac{1}{2} W^{ab}_c\cdot\phi^c dS_{ab}
\end{equation}

Relative to a given choice of background connection
${\stackrel{\circ}{\nabla}}$, equation (13) can be interpreted to say that
the total energy, $M$, of the spacetime is the sum of the energy outside
the horizon (the first term on the right hand side of (13)) and the energy
of the horizon itself (the second term on the right hand side of (13)).  The
terms
in equation (14) can be interpreted in a similar way.  Equations (13) and
(14) will provide the second key ingredient for our proof.

 From the lemma in the appendix and lemma 3.1 of \cite{schutz}, we know that
one can construct a field
configuration ${g_{ab}+ \Delta g_{ab}}$ in a
neighborhood of $\Sigma_f$
such that:\\
1. its mass as given by equation (13) is still $M+\delta M$
and it's angular momentum as given by equation (14) is still $J+\delta J$, but
asymptotically it agrees exactly with  $g_{ab}$.\\
{2.} near the horizon on $\Sigma_f$ , $g_{ab} +\Delta g_{ab}$ agrees
exactly with $g_{ab}+\delta g_{ab}$.

\noindent (Notice that $g+\Delta g$ then has to
violate the field equations in some region far from the horizon.)

Using these facts we can construct the new variation $\Delta g$ in $\Omega$
under which we choose to evaluate equation (9).  $\Delta g$ is such that:\\
{[i]} $\Delta g=0$ at $\Sigma_{i}$\\
{[ii]}
  $\Delta g$ agrees with 1. and 2. above; i.e. ${g+\Delta g}$ has mass
  $M+\delta M$ and angular momentum $J+\delta J$
  on $\Sigma_f$, and $\Delta g$ near $\Sigma_f$ vanishes near spatial
  infinity.\\
{[iii]}
 $\Delta g$ within $\Omega$ vanishes in a
 neighborhood of spatial infinity. In other words $\Delta g$ is of compact
 support within $\Omega$.
 Also, $H$ remains a null surface for the metric $g_{ab}+\Delta
 g_{ab}$, and $\Delta g_{ab}=\delta g_{ab}$ near $H$.

 {}From [iii] just above and (d) of section III.1, we have that on $H$
\begin{equation}
\Delta k^{a} =0
\end{equation}
{}From this property of the variation in conjunction with its other
properties one observes the following two useful facts:
 Since $H$ is still a null surface its null normals must be
proportional to $g_{ab}k^{b}=k_{a}$. So
\begin{equation}
\Delta k_{a}= ck_{a}
\end{equation}
for some real $c$.  And since $\xi$ is null on $H$,
\begin{equation}
\xi^{a}= \alpha k^{a}
\end{equation}
for some coefficient of proportionality $\alpha$.

\noindent To summarize: The trick is to replace $\delta g$ by $\Delta g$.
This allows the variation to be of compact support, freeing us from having
to evaluate variations near spatial infinity.

As a final remark, we note that the variations $\delta$ and $\Delta$  are
identical near $H$ and we shall continue to denote by $\delta$, the
variations of quantities near $H$.

\section {The proof for gravity alone}

We would like to emphasize that what we have described in the preceding
sections  is just the necessary background to do the actual proof of the
first law. Once this background is assumed and absorbed, the proof itself
is extremely simple (cf. \cite{winnipeg}).

 The extremum identity, equation (9), written for the $\Delta$
variation is \\
$\Delta \int _{\partial \Omega} T^{a}_{b} \cdot \xi^{b} dS_{a}=0$.
But $\Delta g=0$ near spatial infinity and on $\Sigma_i$, whence
\begin{equation}
             \Delta\int _{\Sigma_{f} } T^{a}_{b} \cdot \xi^{b} dS_{a}
       \; + \; \delta\int _{H} T^{a}_{b} \cdot \xi^{b} dS_{a} \;=\;0
\end{equation}
Taking the variation of equations (13) and (14) with $\Sigma=\Sigma_f$ we get
\begin{equation}
 -\delta M\;=\; \Delta \int_{\Sigma_{f}} T^{a}_{b} \cdot t^{b}dS_{a}
     \;+\; \delta \int_{S^{2}_{\Sigma_{f} \cap H}} \frac{1}{2}
  W^{ab}_c\cdot t^c dS_{ab}
\end{equation}
and
\begin{equation}
   \delta J\;=\; \Delta \int_{\Sigma_{f}} T^{a}_{b} \cdot {\phi}^{b}dS_{a}
     \;+\;
    \delta \int_{S^{2}_{\Sigma_{f} \cap H}}
       \frac{1}{2} W^{ab}_c\cdot\phi^c dS_{ab}
\end{equation}
Equations (10), (18), (19) and (20) yield then
\begin{equation}
  -\delta M \; + \; \omega\delta J
  \;=\;
   \delta \int_{S^{2}_{\Sigma_{f} \cap H}} \frac{1}{2} W^{ab}_c\cdot\xi^c
    dS_{ab}
   \; - \; \delta \int_{H} T^{a}_{b} \cdot \xi^{b}dS_{a}
\end{equation}
Using the definition of the Noether operator for gravity and Stokes
theorem (and recalling that $\d g=0$ on $\Sigma_i\cap H$) we get
\begin{equation}
   -\delta M\;+\;\omega \delta J
    \; = \;
       \delta \int_{H} G^{a}_{b} \xi^{b} dS_{a}
 \end{equation}
a relation of some interest in its own right.  Notice that (22) can be read as
equating the change in energy or angular momentum to what would be the
corresponding ``energy flux" or ``angular momentum flux" through the
horizon if $\delta G_{ab}$ were interpreted as the stress energy of some
fictitious matter field.

We also note that this relation shows in another way that the left hand
side of (22) must reduce to a surface integral on $\Sigma_f \cap H$,
because on the right hand side we can confine $\delta$ to a neighborhood of
$\Sigma_f\cap H$ on $H$, as in [1], and in the limit can put
$\delta{}g_{ab}\propto\theta(\lambda-\lambda_{f})$, where $\theta$ is the
step function.  Yet another way to show this, perhaps the most systematic
of all, would be to use the theorem in \cite{ontimunu} to derive a
``potential'' for $\d(G^a_b\xi^b)$ from the identity
$\partial_a\d(\sqrt{-g}G^a_b\xi^b)\ideq 0$, see the discussion in
\cite{winnipeg}.

   In evaluating the right hand side of (22), let $\lambda$ be an affine
parameter for the null geodesic generators of $H$ so that
$k^{a}=dx^{a}/d\lambda$.  Then
\begin{equation}
        dS_{a} = - k_{a}d\lambda d^{2}A
\end{equation}
where $d^{2}A$ is the area element of the 2-sphere cross sections of $H$
(and $\lambda$ increases toward the future).  Moreover, by differentiating
equation (17), it follows immediately (in view of the fact that $k^a$ is
the tangent to an affinely parameterized null geodesic, whence $k^a\nabla_a
k^b=0$) that
\begin{equation}
                 d\alpha/d\lambda=\kappa     \label{E24}
\end{equation}
where $\kappa$ is the surface gravity of the stationary black hole, defined
by
\begin{equation}
     \xi^a \nabla_a \xi^b = \kappa \xi^b.
\end{equation}

In the extremum identity, $\xi$ is not to be varied, i.e. $\Delta\xi=0$.
Then since $\delta k^{a}=0$ as well, the right hand side of (22) becomes
\begin{equation}
- \int_{H} \alpha k^{a} k^{b} (\delta (G_{ab}d^{2}A))d\lambda
\end{equation}
Using $G_{ab}=0$ for the stationary solution and the fact that $k^a$ remains
null under the variation, we get
\begin{equation}
    - \oint d^{2}A \int d\lambda\; \alpha k^{a} k^{b}(\delta R_{ab})
\end{equation}

To evaluate $(\delta R_{ab})k^{a}k^{b}$ we use the Raychaudhuri equation
for the rate of expansion of a congruence of null geodesics (see for
example \cite{waldbook}):
\begin{equation}
  d\theta/d\lambda= - (1/2) \theta^{2}
   -\sigma_{ab}\sigma^{ab}+\omega_{ab}\omega^{ab}-R_{ab}k^{a}k^{b}
\end{equation}
where $\theta$ is the expansion, $\sigma$  the shear and  $\omega$  the
twist of the null geodesic congruence with $k^{a}$ parameterized by affine
parameter $\lambda$.
We apply  this equation to the null geodesic congruence on $H$ (for which
anyway $\omega=0$) and take
its $\delta$ variation. Using the fact that the expansion and shear
of the stationary solution vanish we get:
\begin{equation}
   d(\delta\theta)/d\lambda=-(\delta R_{ab})k^{a}k^{b}  \label{E29}
\end{equation}
Inserting this into (27) yields
\begin{eqnarray}
         \delta \int_{H} G^{a}_{b} \xi^{b} dS_{a}
          & = & \oint d^{2}A \int d\lambda \alpha d(\delta \theta)/d\lambda \\
          & = &  (\oint d^{2}A \alpha \delta \theta)|^{f}_{i}
        - \oint d^{2}A \int d\lambda {d\alpha \over d\lambda} \delta \theta
\end{eqnarray}
where $i,f$ refer to the 2-spheres which are the intersections of $H$ with
$\Sigma_{i}$ and $\Sigma_{f}$ respectively.

  But $\delta\theta_{i}=0$ by construction, and $\delta\theta_{f}=0$ under
the assumptions of Lemma 1 (which is given at the end of this section).
Also, $\theta=(1/d^{2}A)\;d(d^{2}A)/d\lambda$ by definition.  Hence,
\begin{equation}
      \delta \int_{H} G^{a}_{b} \xi^{b} dS_{a}
       =   - \oint d^{2}A \int d\lambda
     {d\alpha\over d\lambda} \,
        \delta [ {1 \over d^{2}A} \; {d (d^{2}A) \over d\lambda} ]
\end{equation}
Using (\ref{E24}) together with the fact that for the stationary solution
$d(d^{2}A)/d\lambda=0$, we get for (32)
\begin{equation}
   - \oint d^{2}A \int d\lambda \, \kappa \,
      {1 \over d^{2}A} \, \delta {d(d^{2}A) \over d\lambda}
\end{equation}
\noindent But $\delta\lambda=0$, so
\begin{eqnarray}
       - \delta \int_{H} G^{a}_{b} \xi^{b} dS_{a}
     & = & \kappa \oint\int d\lambda  {d (\delta d^{2}A) \over d\lambda} \\
     & = & \kappa \, \delta (\oint d^{2}A)|^{f}_{i} \\
     & = & \kappa \, \delta (\oint d^{2}A)|_{f}
\end{eqnarray}
where in the last line we have used that the variation $\delta$ vanishes
on $\Sigma_{i}$.  Hence we get by (22)
\begin{equation}
\delta M\;\;=\;\;\kappa \delta A\;+\;\omega\delta J
\end{equation}
where everything is evaluated on $\Sigma_{f}$.

To complete our proof, we have to justify setting
$\delta\theta_{f}=0$, which we do via the following lemma.

\noindent
{\bf Lemma 1}:
Let $(g_{ab},\{\phi\})$ be a stationary black hole solution to the field
equations, where $\{\phi\}$ denotes all fields other than the gravitational
field, and let the stress-energy tensor $T^{ab}$ of $\{\phi\}$ satisfy the
positivity condition that $T^{ab}k_{a}k_{b}\geq0$ for every null $k^{a}$.

Consider the spacetime region $\Omega^{\prime}$ to the future of some
spatial slice $\Sigma_{f}$. Denote the part of the horizon in this region
by $H^{\prime}$.  Let $(g_{ab}+\delta g_{ab}, \{\phi +\delta \phi\})$ be an
infinitesimally differing solution, with the relevant part of its horizon
coinciding with $H^{\prime}$.

Then the variation $\delta\theta$ of the expansion evaluated on
$\Sigma_{f}$ vanishes if either of the following (rather weak) stability
conditions holds:\\
\phantom {kuku}
{\bf (i)} It is not the case that arbitrarily small perturbations of
the metric and the other fields can destroy the horizon; or\\
\phantom {kuku}
{\bf (ii)} $\delta \theta$ approaches zero in the distant future along the
horizon.

\noindent{\bf Proof}:

\noindent {\bf (i)} $(g_{ab}+\delta g_{ab}, \{\phi +\delta \phi\})$
is a solution to the linearized field equations.  Therefore
$(g_{ab}-\delta g_{ab}, \{\phi -\delta \phi\})$
 is also a solution to the linearized equations.
Thus, if for the
former solution,
the variation of the expansion on
$H^{\prime}$  is $\delta \theta$ then for the latter solution,
the variation in the expansion will be $-\delta \theta$.

Since $\theta=0$ for the stationary solution, there exists a solution to
the full Einstein equations in a neighborhood of the stationary solution
whose expansion at first order is $\delta \theta$, which, as just
explained, can be arranged to be negative unless it is always zero.  But it
is known that a negative expansion implies a singular evolution of the
horizon.  Namely, by using (``to all orders'') the Raychaudhuri equation
(28), the Einstein equation and the positive energy condition above, one
can conclude that if the expansion $\theta=\theta_0$ is negative at some
$\lambda =\lambda_{0}$ (the geodesics are converging), then a conjugate
point (caustic) will develop at or before some definite future instant
$\lambda_{1}$, where $\lambda_{1}$ depends on $\theta_{0}$.  But the
occurrence of a conjugate point contradicts the definition of the horizon,
because no geodesic generator can remain on the horizon after reaching a
conjugate point.  Thus, the only possibility is that the generator in
question terminates in a singularity, i.e. that the perturbation destroys
the horizon, contradicting assumption (i).

\noindent {\bf (ii)} Applying the variation $\delta$ to the Raychaudhuri
equation and noticing as before that the expansion, shear and twist of the
stationary solution vanish, we get from (\ref{E29})
\begin{equation}
  d(\delta\theta)/d\lambda
  = -(\delta R_{ab})k^{a}k^{b}
  = -\delta T_{ab}k^a k^b,
\end{equation}
in view of the Einstein equation.  But from (28) and the Einstein equations
it follows that for the stationary solution
\[
    R_{ab}k^a k^b=T_{ab}k^a k^b=0
\]
Hence the positivity condition on the stress energy tensor implies that
\[
    \delta T_{ab} k^a k^b \geq 0
\]
for arbitrary variations of the fields.  But as
$\delta T_{ab}$ is linear in the infinitesimal variations
$(\delta g_{ab},\{\delta\phi\})$,
this is impossible unless $\delta T_{ab} k^a k^b =0$ for  all  variations,
whence
\begin{equation}
d(\delta\theta)/d\lambda=0
\end{equation}
for all variations.
Hence $\delta \theta_{f}= \delta \theta_{\infty} $ where
$\delta \theta_{\infty}$ is $\delta \theta$ evaluated at large
affine parameter and $\delta\theta_{f}$ is $\delta \theta$
evaluated at $\Sigma_{f}$.  But $\delta \theta_{\infty}=0$ by
 assumption.  Hence $\delta \theta_{f}$ vanishes as well.
(Notice that if $\Sigma_f$ intersected the horizon in a bifurcation surface
as in \cite{wald} then, since by definition $\alpha=0$ at such a surface,
$\alpha \delta \theta_{f}=0$
would be immediate and lemma 1 would not be needed.)

\section{The proof for gravito-electric spacetimes}

\subsection{The Electromagnetic Noether operator}

In order to generalize (37) to gravito-electric spacetimes, we will
apply the
extremum identity (9) to field variations about a stationary, charged black
hole solution of the Einstein-Maxwell equations.
 From the general argument in section 1, we know that
$\delta M$ and $\delta J$
will be expressible entirely in terms of ``horizon
variables", and since the total Action is now the sum of $S_{grav}$ and
a single electromagnetic term, we know that a single extra term will
appear in (37). We can find this term by using the gravito-electric versions
of equations  (9), (13) and (14).  To do this, we
need  expressions for the relevant Noether operators. We display them below
(for details see \cite{chandra}).

The Action for the electromagnetic field is $S_{em}=(-1/4)\int F_{ab}F^{ab}dV$
 where
$F_{ab}=\nabla_{a}A_{b}-\nabla_{b}A_{a}$ is the electromagnetic field tensor
and $A_{b}$ its potential.  The Action for the combined gravitational
and electromagnetic fields is $S_{grav}+S_{em}$ ($S_{grav}$ has been
defined in section II).
The action of the gravito-electric Noether operator on a vector field
$\xi^b$ is:
\begin{equation}
     {\cal T}_{total\;b}^{a} \cdot \xi^{b}
        := \sqrt{-g} \; T_{total\;b}^{a} \cdot \xi^{b}
         = \sqrt{-g} \; (T_{em\;b}^{a} +T_{grav\;b}^{a}) \cdot \xi^{b}
   \label{E40}
\end{equation}
where $T_{grav\;b}^{a}\cdot\xi^{b}$ is given in equation (3) and
${T_{em}}^{a}_{b}$  is given \cite{chandra} by
\begin{eqnarray}
  T_{em\;b}^{a} \cdot \xi^{b}
    & = & F^{ab}{\cal L}_{\xi}A_{b} - (1/4)F_{cd}F^{cd} \xi^{a} \\
    & = & T_{S\;b}^{a}\xi^{b}-\nabla_{b}(\xi^{c}A_{c})F^{ab},  \label{E41}
\end{eqnarray}
$T_{S}^{ab}$ being the stress energy tensor of the electromagnetic field,
\begin{equation}
T_{S}^{ab}=F^{ac}F^{b}_{\phantom{B} c}-(1/4)g^{ab}F_{cd}F^{cd}
\end{equation}
The electromagnetic analogue of (3) is therefore
\begin{equation}
  T_{em\;b}^{a} \cdot \xi^{b}
  = (T_{S\;b}^{a}+(\nabla_{c}F^{ac})A_{b}) \xi^{b}
    - \nabla_{b} (F^{ab}A_{c}\xi^{c})
\end{equation}
and the total Noether current is---as in (3)---the sum of a divergence
with a term which vanishes ``on shell":
\begin{eqnarray}
   T^{a}_{total\;b} \cdot \xi^{b}
   & = &
     [ T_{S\;b}^{a} - G^{a}_{b} + (\nabla_{c}F^{ac})A_{b} ] \xi^{b}
      \nonumber \\
   & \; &
   + \nabla_{b} (W^{ab}_c\cdot\xi^c - F^{ab}A_{c}\xi^{c})
\label{GENoether}
\end{eqnarray}
with $W^{ab}_c\cdot\xi^c$ given by (3).

\subsection{The extremum identity}

The  extremum identity reads now
\begin{equation}
  \delta\int _{\partial \Omega} T^{a}_{total\;b} \cdot \xi^{b} dS_{a} \;=\;0
\end{equation}
As before, we will apply this with $\xi^{a}=t^{a}+\omega \phi^{a}$ where
$t^{a}$ and $\phi^{a}$ are
Killing fields  which are asymptotically time translational and rotational
respectively; $\xi^a$ is that killing field which is null
 on the horizon and $\omega$ is the angular velocity of the black hole.

\subsection{Some useful properties of the stationary solution}

The variations will be made around a stationary black hole configuration,
$g_{ab},\;A_{c}$. The following properties of this stationary black hole
solution will be of use:\\
\phantom{kuku}
  (a) ${\cal L}_{\xi}A_{a}=0$. (This amounts to a suitable gauge
  choice and
  was implicitly assumed when we wrote down the extremum identity above)\\
\phantom{kuku}
(b) Claim: On the horizon $F^{ab}\xi_{a}=\gamma \xi^{b}$ for some
  function $\gamma$ . \\
\noindent {}{}{}{}Proof: Using (28) for  the stationary solution on the
horizon,
the Einstein equations and the proportionality of the null geodesic
 horizon-generators to the Killing field $\xi$, one gets
\[ T_{Sab}\xi^a \xi^b=0\]
on the horizon.  Substituting the explicit form of $T_{Sab}$ from
(43) into this
equation, one
concludes that $F^{ac}\xi_c$ is a null vector. From the antisymmetry of
$F^{ac}$, one concludes that this null vector has to be proportional to
$\xi^a$, since it is both null and orthogonal to $\xi$. \\
\phantom{kuku}
(c)  The `electric potential' $-\xi^{a}A_{a}$ is constant on the
horizon.  (Contract the identity
$\xi^{a}F_{ab}={\cal L}_{\xi}A_{b}-\nabla_{b}(\xi^{a}A_{a})$
with any $v^b$ tangent to the horizon and use (a).)  Following the common
notation, we will write $V =  - \xi^{a}A_{a}$ for this constant.

These relationships express the fact that the horizon behaves
like a conducting surface : in equilibrium (condition a) the
electric potential must be constant (condition c) and hence the
``electric field" must be perpendicular to the surface (condition b)
(see, for example \cite{membrane}).

\subsection{The variation `$\delta$'}

We use the same notation as in previous sections for spacetime regions,
volume elements and anything involving only the metric.

The conditions on `$\delta $ variations' are similar to those  in
section III.1. We choose $( \delta A_{c}, \delta {g}_{ab} )$ as follows:\\
\noindent (a) On $\Sigma_i$ and to it's past, $\delta g_{ab}=\delta A_{c}=0$
(we don't want to involve $\Sigma_i$ at all)\\
\noindent (b) On $\Sigma_f$ and to it's future
$(g_{ab}+ \delta g_{ab},\;A_{c}+\delta A_{c})$
is a black hole solution to the field equations with energy $M+\delta M$
angular momentum $J+\delta J$ and charge $Q+\delta Q$.
We arrange that the horizons of the varied and unvaried solutions coincide. \\
\noindent
(c), (d) These conditions are identical to conditions (c)
and (d) of Section III.1.

We also choose  our background connection, $\stackrel{\circ}{\nabla}$,
as in III.1. As before, we have to
deal with non-solutions in the region between $\Sigma_{i}$ and $\Sigma_{f}$.

\subsection{The variation `$\Delta$'}

We face a similar technical problem as in Section III.2.  To get around it,
we repeat the steps followed there.  We first define expressions for the
energy $M$ and angular momentum $J$ of a field configuration on an
asymptotically flat spatial slice $\Sigma$ which reduce to the correct
values when the source free Einstein-Maxwell equations are satisfied,
but which are also useful for fields which are not solutions
(these expressions follow directly from (\ref{GENoether}) and the fact that
the total energy and angular momentum on a boundary-less slice can always
be written as
$\int_{\Sigma}T_{total\;a}^{b}\cdot  \xi^{a}dS_{b}$ with $\xi$ the
relevant generator), to wit:
\begin{equation}
     -M =
         \int_{\Sigma}T_{total\;a}^{b}\cdot  t^{a}dS_{b}
       + \oint_{S^{2}_{\Sigma \cap H}}\frac{1}{2} F^{ab}A_{c} t^{c}dS_{ab}
       + \oint_{S^{2}_{\Sigma \cap H}}\frac{1}{2} W^{ab}_c\cdot t^c dS_{ab}
\end{equation}
\begin{equation}
   J =
       \int_{\Sigma}T_{total\;a}^{b} \cdot \phi^{a}dS_{b}
     + \oint_{S^{2}_{\Sigma \cap H}} \frac{1}{2} F^{ab}A_{c} \phi^{c}dS_{ab}
     + \oint_{S^{2}_{\Sigma \cap H}} \frac{1}{2} W^{ab}_c\cdot\phi^c dS_{ab}
\end{equation}

Using arguments similar to those in section III.2
 one can introduce variations of compact support of all the fields (i.e. $A$
and $g$) such that    the new field configuration on $\Sigma_{f}$ has
mass $M+\delta M$ and angular momentum $J+\delta J$,
 and is identical with ($A+\delta A, g+\delta g$) near the
horizon. We will not display  the relevant patching arguments since they
are similar to those of Appendix B. We simply note that in order to complete
the patching arguments, one has to make use of the asymptotic conditions on
$A_{c}$ as well.  These conditions are listed in Appendix C.

Let us, as before, call this new variation of compact support the $\Delta$
variation.  Just as earlier, it satisfies the analogs of properties
[i]--[iii] of Section III.2, as well as equations (15)--(17).  And, we can
again continue to denote $\Delta$ on the horizon by $\delta$, as there is
no difference between the two variations there.

\subsection{The proof}

We have the extremum identity:
\begin{equation}
  \Delta \int_{\partial \Omega}{T}_{total\;b}^{a}\cdot\xi^{b}dS_a=0
\end{equation}
\begin{equation}
\Rightarrow \Delta \int_{\Sigma_{f}}{T}_{total\;b}^{a}\cdot\xi^{b}dS_a
         +  \delta \int_{H}{T}_{total\;b}^{a}\cdot\xi^{b}dS_a=0
    \label{E50}
\end{equation}
Using the above expressions for $M$ and $J$ with $\Sigma=\Sigma_f$, and
taking respectively $\xi=t$ and $\xi=\phi$ in (\ref{E50}) yields, in view
of (\ref{E40}),
\begin{displaymath}
  - \ \delta M =
     \delta \int_{S^{2}_{\Sigma_{f} \cap H}}\frac{1}{2} W^{ab}_c\cdot t^c
dS_{ab} \;
    - \; \delta \int_{H} T^{a}_{grav\;b}\cdot t^{b}dS_{a}
 \; - \; \delta \int_{H} T^{a}_{em\;b}\cdot t^{b}dS_{a}
\end{displaymath}
\begin{equation}
  \;\;
\;\;\;\; + \; \delta \int_{S^{2}_{\Sigma_{f} \cap H}} \frac{1}{2}
            F^{ab}A_{c} t^{c}dS_{ab}
\end{equation}
\begin{displaymath}
   \delta J
  =
     \delta \int_{S^{2}_{\Sigma_{f} \cap H}}
             \frac{1}{2} W^{ab}_c\cdot\phi^c dS_{ab} \;
    - \; \delta \int_{H} T^{a}_{grav\;b}\cdot\phi^{b} dS_{a}
 \; - \; \delta \int_{H} T^{a}_{em\;b}  \cdot\phi^{b} dS_{a}
\end{displaymath}
\begin{equation}
   \;\;\;\;\;\;+ \; \delta \int_{S^{2}_{\Sigma_{f} \cap H}}
                  \frac{1}{2} F^{ab}A_{c} \phi^{c}dS_{ab}
\end{equation}
Then the same manipulations of the gravitational terms in these equations
as we performed in section IV produce here
\begin{equation}
      - \ \delta M + \omega \delta J = ({\rm grav})\;+\;({\rm em}) \label{E53}
\end{equation}
where
\begin{equation}
  ({\rm grav})= \delta \int_{H} G_{b}^{a}\xi^{b} dS_{a}
              = -\kappa \delta A +\oint_{S} d^{2}A\alpha\delta \theta
\end{equation}
is the same expression as before (with $\xi^a$ again denoting
$t^{a}+\omega\phi^{a}$), and
\begin{equation}
 ({\rm em}) = \delta ( - \int_{H} T^{a}_{em\;b}\cdot\xi^{b}dS_{a}
                       + \frac{1}{2} \oint_{S} F^{ab}A_{c}\xi^{c}dS_{ab})
            \label{E55}
\end{equation}
Here $A$ denotes the area of the 2-sphere $S := \Sigma_{f} \cap H$
and $\oint_S$ denotes integration over this 2-sphere.
Notice that in (\ref{E55}) it is  {\it not} convenient to convert the
integral over $\Sigma_{f} \cap H$ to an integral over $H$ as we did with
the corresponding  integral for gravity in equation (21).

    As expected, our expression for $-\d M+\omega\d J$ is just the same as
before, with the exception of the additional term `(em)' in equation
(\ref{E53}).  To convert this term to a surface integral over $S$ (as we
already know must be possible on general grounds), we proceed as follows.
Direct substitution of eq. (\ref{E41}) into (55) yields
\begin{equation}
  ({\rm em}) = I_1 + I_2 + I_3
\end{equation}
where
\begin{eqnarray}
         I_1 & := -\d \int_H F^{ab} \Lie_\xi A_b dS_a \nonumber\\
         I_2 & :=  \d \int_H \frac{1}{4} F^{cd}F_{cd} \xi^a dS_a \nonumber\\
         I_3 & := \d \oint_S A_c \xi^c \frac{1}{2} F^{ab}dS_{ab}. \nonumber
\end{eqnarray}
Now $I_3$ is already a surface integral over $S$, and $I_2$ is easily seen to
vanish in light of (23), (16) and the fact that $\xi \cdot \xi$ and
$\d\xi^a$ both vanish.  For $I_1$, we can transform its integrand using
(V3.a) to get
\[
   F^{ab}\Lie_\xi\d A_b
    = \Lie_\xi(F^{ab}\d A_b)
    = \nabla_c Y^{ac} - \nabla_c(F^{ab}\d A_b)\xi^a
\]
where $Y^{ac} \ideq F^{ab}\d A_b\xi^c - (a \leftrightarrow c)$.
The final term does not contribute to the integral $I_1$ because
$\xi^a dS_a=0$ by (17) and (23), and the first term yields an integral
which can be converted by Stokes' theorem to obtain
\[
   I_1 = - \, \oint_S F^{ab} \d A_b \xi^c dS_{ac}
\]
(remember that $\d=0$ on $\Sigma_i$).
Returning to $I_3$ for a moment, we can write
\begin{equation}
  I_3 = A_c\xi^c\d\int\half F^{ab}dS_{ab} + \int\half dS_{ab}F^{ab}\d(A_c\xi^c)
  \label{EC}
\end{equation}
where we have used (V.3c) to take $-V=A\cdot\xi$ out from under the
integral sign.  It is then a matter of a few lines of straightforward
algebra to show that the second integral in (\ref{EC}) cancels with $I_1$.
(The algebra uses (23), (V.3b), and the fact that the surface element
$dS_{ab}$ of $S$ can be written as
\begin{equation}
  dS_{ab} = d^2A \, (k_a l_b - k_b l_a)
\end{equation}
where $l^a$ is any future-null vector orthogonal to $S$
such that $k_a l^a=-1$.)  Thus we are
left with
\[
  ({\rm em}) = I_1+I_2+I_3 = A_c\xi^c\d\oint_S\half F^{ab}dS_{ab}
\]
or
\begin{equation}
    ({\rm em}) = - \ V \delta Q
\end{equation}
because of the perfectly general relation
\begin{equation}
        Q = \oint_S\half F^{ab}dS_{ab}
\end{equation}
$Q$ being the charge of the black hole, evaluated as an (outward) flux
through $S=H\cap\Sigma_f$.

Substituting this back into equation (\ref{E53}), we get
\begin{equation}
   -\delta M + \omega\delta J
    =
  -\kappa\delta A + \oint_{S}d^{2}A\alpha\delta\theta \;-\; V\delta Q
  \label{almost}
\end{equation}
Finally, Lemma 1 comes to our aid to tell us once again that
$\delta\theta=0$ on $S$, so, putting this back in equation (\ref{almost}),
we get our generalized extremality theorem for the gravito-electric case:
\begin{equation}
     \delta M \; = \; \kappa\delta A + \omega\delta J \;+\; V\delta Q
\end{equation}

\section{Conclusions}
As we intimated in the Introduction, the primary impetus for writing this
paper came from our curiosity about how the extremality theorem of
reference \cite{schutz} would be modified by adaptation to the spacetime
region exterior to a black hole.  Here we would like to comment more at
length on the wider significance of the results obtained, and more
generally on that of the first law itself.  In the course of these comments
we also will mention some further related work which either has been done
already, or would seem to be worth doing.

As its name makes clear, the most important implications of the so-called
first law pertain to the thermodynamic attributes of black holes; however
the associated issue of stability is important in its own right  (and
arises already in the purely classical setting).

By virtue of being time-independent, a stationary black hole solution may
be characterized as a state of equilibrium of a self-gravitating system.
If this equilibrium is to be stable in the thermodynamic sense, then of
course it must (at least locally) be a maximum of the entropy $S$ (we use
`$S$' for entropy in this section).  Conversely, if some configuration does
maximize $S$, then the second law of thermodynamics implies that the
equilibrium state in question is indeed a stable one.  Assuming that $S$ is
a smooth function on the relevant space of ``configurations'' or
``states'', a more or less necessary and sufficient condition for
equilibrium is thus that $S$ have vanishing gradient and negative definite
Hessian.  More generally, one may just define an equilibrium configuration
to be an extremum of $S$, and diagnose its (thermodynamic or ``secular'')
stability by the behavior of the Hessian there.

Actually, the criteria we have just stated need to be qualified, as
equilibria need not be unconditional maxima of $S$, but only so at fixed
values of the relevant conserved quantities.  In the black hole situation,
the relevant quantities will usually be the energy $E$, the angular
momentum $J$, and the electric charge $Q$.  If we identify the area $A$
with the entropy (up to a numerical factor), then a necessary condition for
a black hole configuration to represent a thermodynamic equilibrium state
is that a relation of the form (62)
(i.e. $dA = \beta dE - \beta \omega dJ - \beta V dQ$) hold for {\it
arbitrary} variations of the fields.  (Clearly this equation entails
extremality of $S$ at fixed $E$, $J$ and $Q$; and conversely, extremality
implies a relation of this general form, although of course, it does not
give us the specific expressions (24), (V.3c), etc. for the coefficients
$\beta$, $V$, $\omega$, etc. which appear in (62).)  This relation to
thermodynamic equilibrium is the main reason why it is important that (62)
hold more generally than just for variations from one stationary solution
to another.

Although the satisfaction of equation (62) qualifies a black hole solution
as an equilibrium state in the above sense, it has nothing directly to say
about stability.  As we just pointed out, this concerns the second
derivatives of the entropy, or in the classical limit, of the horizon
area.  In thus using $A$ as the measure of classical stability, we of
course presuppose the more general identification of area with entropy,
which rests first of all on (62) itself.  In the $ \hbar \rightarrow 0$
limit, this geometrical contribution dominates all other sources of
entropy, whence the latter can be ignored.  The consequent fact that $A$
can be used to diagnose classical stability is consistent with the
interpretation of the classical law of area increase as expressing the
second law of thermodynamics for black holes, but unfortunately the area
law can be justified as an independent classical fact only to the extent
that ``cosmic censorship'' holds---one of many examples of the close
interconnection between cosmic censorship and black hole thermodynamics.
If we could establish cosmic censorship, or prove the area law on some
independent basis then area maximization could be used as a criterion of
stability independently of any thermodynamic argument: all that matters is
that horizon area be (classically) non-decreasing, and therefore a valid
``Lyapunov functional''.

In any case, in order to investigate stability on the basis of horizon
area, one should determine the negative-definiteness (or lack thereof) of
the ``reduced Hessian'', $d^2A-\beta d^2E$, or more generally of
$d^2A - \beta d^2E - \beta V d^2 Q - \beta \omega d^2 J$.  One might hope
to prove negative-definiteness in all directions except those corresponding to
``super-radiant modes'' in the rotating case.  We have not studied this
Hessian directly; however there exists an indirect way to investigate its
eigensigns, namely the so-called turning point method commonly employed in
studies of stellar stability and the stability of other astrophysical
objects (cf.\cite{instabcrit,stabcrit,axistab}).  This method can prove
instability, but can only make stability plausible.  As applied to the
Reissner-Nordstrom and Kerr-Newman families of black holes, it indicates
stability, or rather it indicates that nowhere along the sequence does
a new unstable mode come into being and remain for a finite range of
parameter value \cite{stabcrit}.  Indirectly, this implies that the above
reduced Hessian is indeed negative in the required sense.  A direct
confirmation would be of considerable interest.

We believe that our method of proving the ``extended first law'' for
Einstein gravity is simpler than previous ones (cf. \cite{winnipeg}), and
that for this reason, it clarifies as well the origin of the ``unextended
first law'', the one applying only to variations from one stationary
solution to another.  Aside from this possibly greater simplicity and the
associated thoroughgoing spacetime character of the definitions and the
derivation, the most important way in which our result differs technically
from earlier ones is that (without making any reference to a possible
``bifurcation surface'') it allows the hypersurface $\Sigma$ on which the
energy, area, etc. are evaluated to intersect the horizon in an arbitrary
manner, and in particular to intersect it arbitrarily far into the future.
(The importance of this type of hypersurface freedom was stressed also in
\cite{tedetal}.) This freedom in the choice of $\Sigma$ is mandated
physically, since it is only some future-segment of a Schwarzschild metric
(say) that is relevant to an astrophysically realistic black hole like one
formed in stellar collapse.  But the more important issue of principle, in
our view, has to do with the second law itself.

For the theory of black hole thermodynamics to achieve a satisfactory
status, it will be necessary not only to derive the equilibrium value of
the entropy from first principles, but also to explain why the law of
entropy increase continues to hold when black holes are present.  One
attempt to foresee how such a proof would go \cite{qbh}, presupposes that
(say in a setting which is near enough to classical that an approximately
well defined identification of individual spacelike hypersurfaces can be
made) the total entropy in question is that of an effective quantum
density-operator associated to the portion of each such hypersurface lying
outside the black hole(s).  The second law is then identified with the
assertion that this entropy is weakly increasing as the hypersurface
advances in time along the horizon (or at spatial infinity).  For
consistency, then, it would be necessary that our earlier considerations on
area extremization hold for such hypersurfaces, as well as for ones tied,
for example, to a bifurcation 2-surface.  In fact, the desire to prove
extremality for such hypersurfaces was our second main reason for writing
this paper.

Extremal theorems can also be approached from a Hamiltonian point of view,
as we mentioned in the Introduction.  Recently, such an approach was
employed in \cite{wald}, the result being a theorem similar to ours, except
that the hypersurface on which the canonical variables were defined was
assumed to meet the horizon in a bifurcation 2-surface.  Other recent
papers dealing with these same general issues are
\cite{tedmyers,tedkangmyers,waldnoether},
the last two of which are rather similar in spirit to our own.
Some of these are concerned with higher derivative or higher dimensional
gravity. There is little about our methods which would not seem
to apply in those cases as well, but that is a question we have not
looked into.

In concluding, let us mention a final technical point and a final issue for
further study.  In our derivation we have defined the energy and angular
momentum directly from the Noether operator expressions described in
Section II.  For completeness, one would like to confirm that these
definitions agree in general with the ADM ones, given appropriate
asymptotic falloffs.  This has been done and will be reported by one of us
in another place \cite{mad}.  Related to this way of introducing conserved
quantities, is the fact that $E$, for example, splits naturally into a sum
of an ``exterior energy''---a spatial volume integral---and a surface
contribution from the horizon (see equations (13) and (47), or (14) and
(48) for $J$).  This raises the question whether (perhaps via a suitable
choice of background connection $\stackrel{\circ}{\nabla}$) these separate
terms (or their variations)
might have any individual significance, so that one could
in this way meaningfully disentangle the energy or angular momentum ``of
the black hole'' from that ``of the exterior matter''.

\vspace{2mm}

The work of RDS was partially supported by NSF grant PHY 9307570.

\vspace{2mm}

\appendix{\Large\bf Appendix}
\section{The asymptotic fall off conditions on the metric}

The following notations/conventions will be used in the rest of the Appendix.
Fix an asymptotically Minkowskian coordinate system $(t,x,y,z)$ in a
neighborhood of spatial infinity. Greek indices will denote
 4-d spacetime Cartesian components, Latin indices $i,j,k...$ will denote
 3-d spatial Cartesian components and Latin letters $a,b,c...$ will be
 abstract indices.
The matrix  $\eta_{\mu\nu}$ representing the fixed flat metric at spatial
infinity will be diagonal with components ($-1$,1,1,1).
We also choose our fixed background connection $\stackrel{\circ}{\nabla_{\mu}}$
to coincide with the Cartesian coordinate derivative $\partial_{\mu}$ in a
neighborhood of spatial infinity.

 We define the inversion map, ${\cal I}$,
on a neighborhood of spatial infinity in the following way:
\begin{equation}
                 {\cal I}(x^{\mu})= -x^{\mu}
\end{equation}
The point-map $\cal I$ naturally induces a map on tensor fields in a
neighborhood of spatial
infinity which we shall also call ${\cal I}$.
Our falloff conditions will entail that, to leading nontrivial order,
\begin{eqnarray}
              {\cal I}g_{ab} & = & g_{ab}\nonumber \\
              {\cal I}{\stackrel{\circ}{\nabla}}_{c} g_{ab} & = &
                   {\stackrel{\circ}{\nabla}}_{c} g_{ab}
\label{E64}
\end{eqnarray}

The full set fall off conditions on the spacetime metric $g_{ab}$ at spatial
infinity in terms of $r={(x^{2}+y^{2}+z^{2})}^{1/2}$ are as follows
(spatial infinity being approached as $r\rightarrow\infty$
at fixed $t$).
\\
\noindent Let $h_{\mu\nu}:=g_{\mu\nu}-\eta_{\mu\nu}$.  Then \\
\noindent
(a)  $h_{\mu\nu} =\; \alpha_{\mu\nu}(x^{\tau})/r\;+\;O(1/r^{2})$,\\
  where $\alpha_{\mu\nu}(x^{\tau})$ is bounded and
  $\alpha_{\mu\nu}(-x^{\tau})=\alpha_{\mu\nu}(x^{\tau})$\\
\noindent
(b) $\partial_{\alpha}h_{\mu\nu} =
    \beta_{\alpha\mu\nu}(x^{\tau})/r^{2}+O(1/r^{3})$,\\
  where $\beta_{\alpha\mu\nu}(x^{\tau})$ is bounded and
    $\beta_{\alpha\mu\nu}(-x^{\tau})=-\beta_{\alpha\mu\nu}(x^{\tau})$\\
\noindent
(c) $\partial_{\alpha}\partial_{\beta}h_{\mu\nu}=O(1/r^{3})$

\vspace{3mm}

Now consider an asymptotically flat spatial slice, $\Sigma$, which is
asymptotically a $t=a$ slice ($a$ being some constant). Conditions (a) and
(b) above relate fields on the $t=a$ slice to those on the $t=-a$ slice,
but for our estimates, we would like conditions dealing exclusively with
fields on a single slice. The fields at $(x^{i}, t=-a)$ are related to
those at $(x^{i}, t=a)$ by an integration of their derivatives over a
finite time interval $(\Delta t=2a)$. Using this fact we obtain, on a fixed
slice  $\Sigma$:\\
\noindent
(a1)  $h_{\mu\nu} =\; \alpha_{\mu\nu}(x^{i})/r\;+\;O(1/r^{2})$\\
\noindent
(b1) $\partial_{\alpha} h_{\mu\nu} = \beta_{\alpha\mu\nu}(x^i)/r^{2}
                                    + O(1/r^{3})\\$
where $\alpha_{\mu\nu}$ and $\beta_{\alpha\mu\nu}$ in (a) and (b) have been
restricted to $\Sigma$ and are respectively even and odd functions
of $x^i$ for large $r$.

The evenness conditions expressed in (\ref{E64}), (a), (b), (a1) and (b1)
are often called ``parity conditions''.  In this appendix, we will employ
the term more generally to denote our falloff conditions as a whole.

\section{Asymptotic patching and angular momentum}

{\bf Lemma:}
Let $g_{ab}$ and $\hat{g}_{ab}$ be two metrics which satisfy the parity
conditions of part A of this appendix, and set
  $J(g_{ab}):=\int_{\Sigma}T_{a}^{b}\cdot\phi^{a}dS_{b}$
where $T^a_b$ is the gravitational Noether operator and
$\phi^a$ is an asymptotically Killing, rotational vector field which
commutes with $\grad0$ in a neighborhood of infinity.
In a neighborhood of  $\Sigma$, let
$$
  \bar{g}_{ab}:=\;bg_{ab}+(1-b)\hat{g}_{ab}
$$
where the ``patching function''  $b(x)$ is defined as follows.
Fix a smooth function $f$ on $[0,\infty]$ such that $f(y)=1$ for $y<1$ and
$f(y)=0$ for $y>2$.  Choose a radius $R$ large enough so that the
asymptotic coordinate system is defined for $r>R/2$, and set
$b(x)=f(r/R)$.  (Strictly speaking, $r$ might not be defined throughout the
region enclosed by the $r=R/2$ two-sphere, but we still put $b\ideq 1$ on
that region.)  Then the difference $J(\bar{g}_{ab})-J(g_{ab})$ can be made
as small as desired by choosing $R$ big enough.

\vspace{5mm}

\noindent
{\bf Proof}:
A useful expression for $T_{\mu}^{\nu}\cdot \phi^{\mu}$ is:
\begin{eqnarray}
 T_{\sigma}^{\tau}\cdot \phi^{\sigma}
  & = & - \frac{1}{2} (\nabla_{\eta}\phi_{\rho})
    [2\Gamma_{\sigma\alpha}^{\tau}g^{\eta\alpha} g^{\sigma\rho}
    -\Gamma_{\nu\sigma}^{\nu}
    (g^{\eta\tau} g^{\sigma\rho}+g^{\eta\sigma} g^{\tau\rho})
    +\Gamma_{\nu\sigma}^{\nu}g^{\eta\rho} g^{\sigma\tau}
   -\Gamma_{\beta\alpha}^{\tau}g^{\beta\alpha} g^{\eta\rho}] \nonumber\\
   &  & + \frac{1}{2} \phi^{\tau}[R-\nabla_{\rho}
 (\Gamma_{\beta\alpha}^{\rho}g^{\beta\alpha}
 -\Gamma_{\nu\sigma}^{\nu}g^{\sigma\rho})]
  \label{E65}
\end{eqnarray}
(In this equation, `$R$' stands for the Ricci scalar, of course; it is not
the radius parameter involved in the definition of the patching function.)
Since the second derivatives $\partial\partial g$ cancel out in the second
term of (\ref{E65}), the main term to worry about is the first, which might in
principle decay as slowly as $O(1/r^2)$ because $\partial\phi$ is $O(1)$
and $\Gamma$ is $O(1/r^2)$.  Thus the crucial fact about the first term is
that the coefficient of $\nabla_\eta\phi_\rho$ is symmetric in $\eta$ and
$\rho$.  Using this, and taking $R$ big enough so that $\phi^\sigma$ is a
%
%
%
Killing vector for the background flat metric $\eta_{\mu\nu}$ for $r>R$ one
has symbolically (in the spirit of \cite{schutz}),
\begin{equation}
  I = J(\bar g)-J(g) =
                 \int_{r>R}\partial\bar{g}\partial\bar{g}\phi\;+\;
                 \int_{r>R} \partial \phi (\bar{g}-\eta)\partial\bar{g}
\end{equation}
As a result, the contributions to $I$ are of the following type
(as could also have been seen from the footnote to
equation (4) in reference \cite{chandra}):\\
\noindent
$I_{1}=\int_{R}^{2R} bb (\partial h) (\partial h) \phi\;\;$
$I_{2}=\int_{R}^{2R} b (\partial b) (\partial h) h \phi\;\;$
$I_{3}=\int_{R}^{2R} hh (\partial b) (\partial b) \phi \;\;$ \\
\noindent
$I_{4}=\int_{R}^{2R} (\partial \phi)hbb \partial h\;\;$
$I_{5}=\int_{R}^{2R} (\partial \phi)hbh \partial b \;\;$
$I_{6}=\int_{R}^{\infty} [(\partial h) (\partial h) \phi+
(\partial \phi )h \partial h]$\\
\noindent
$I_{7}=\int_{2R}^{\infty} [(\partial h) (\partial h) \phi
+ (\partial \phi )h \partial h]$\\
(Note: In the above all indices have been suppressed and by $h$ we mean either
$h$ or $\hat{h}$)

Now since $\partial b = O(1/r)$, none of these integrals can be worse than
logarithmically divergent.  In fact we now show that they all are bounded
and vanish as $R\rightarrow\infty$.  We use heavily that $b$ and the
$\alpha_{\mu\nu}$ are even functions on the 2-sphere at constant $r$ and
that the $\partial_\mu b, \beta_{\alpha\mu\nu}$ and $\phi^\mu$ are odd
functions on the 2-sphere at constant $r$.  We denote the element of solid
angle on the 2-sphere by $d\Omega$.  Consider $I_1$ as a typical example.
We have
\begin{eqnarray}
  I_{1} & = & \int_{R}^{2R}b\, b\, \partial h\, \partial h \, \phi\nonumber \\
        & \leq &  \int_{R}^{2R}bb(\beta/r^{2})(\beta/r^{2}) \phi r^{2}drd\Omega
  \;+\;
       \int_{R}^{2R}(bb C_{1}/r^{5}) \phi r^{2}dr d \Omega \nonumber
\end{eqnarray}
where for ease of writing we have omitted absolute value signs around the
the members of the inequality.
Using the parity conditions for $b, \beta, \phi$ and that
$|\phi^{\mu}|<2R,\; |b|<1$ we
get:
\begin{equation}
I_{1}\leq \int_{R}^{2R}(odd\; function/r^{3})r^{2}dr d\Omega\;+\;
         8\pi C_{1} /R
\end{equation}
$\Rightarrow I_{1}\leq A_{1}/R$ for some constant $A_{1}$.  In a similar
fashion, $I_{i}$ for $i=2$ to $5$ are all bounded by some $A_{i}/R$.  (To
show this, one also has to use, in addition to parity arguments, the fact
that $|\partial b| \leq constant/R$, which follows directly from its
definition as $f(r/R)$.)
Using parity arguments it is similarly easy to show that $I_{6},\;I_{7}$ are
also bounded by some $A_{6}/R$ and $A_{7}/R$  respectively.
Therefore, as $R \rightarrow \infty$ all the above contributions to $I$
vanish, and we have proved the lemma.

Remark: In relation to the case of a slice without boundary, we could also
apply this lemma to a proof of an extremum theorem for angular momentum
along the lines of the extremum theorem for mass in \cite{schutz}
(cf. \cite{winnipeg}).   Notice in this connection that, our asymptotic
conditions are about as weak as one could expect to be physically relevant,
and as such are less restrictive than those used in the proof of the
extremum theorem for angular momentum given in \cite{nahmad}.

\section{The asymptotic fall off conditions on $A_{a}$}

We use the notation of section A above.  As there, our fall off conditions
will involve parity conditions under the inversion map ${\cal I}$,
requiring to leading order in $(1/r)$ that
\begin{equation}
                 {\cal I} A_{a} = - A_{a}
\end{equation}
More specifically our conditions are

\noindent
  (a)  $A_{\mu} = \; \chi_\mu(x^\tau) / r \; + \; O(1/r^2)$
       where $\chi_{\mu}(x^{\tau})$ is bounded and
       $\chi_{\mu}(-x^{\tau})=\chi_{\mu}(x^{\tau})$

\noindent
  (b) $\partial_{\nu} A_{\mu}=\rho_{\mu\nu}(x^\tau)/r^2 + O(1/r^{3})$
      where $\rho_{\mu\nu}$ is bounded and
              $\rho_{\mu\nu}(-x^{\tau})=-\rho_{\mu\nu}(x^{\tau})$

\noindent (c) $\partial_{\alpha}\partial_{\beta}A_{\mu}=O(1/r^{3})$

 The same reasoning as in A shows that
on a `$t=$constant' slice $\Sigma$,  (a) and (b) imply:\\
\noindent (a1)  $A_{\mu}
=\; \chi_{\mu}(x^{i})/r\;+\;O(1/r^{2})$\\
\noindent (b1) $\partial_{\nu} A_{\mu}=\rho_{\mu\nu}(x^{i})/r^{2}
                 +O(1/r^{3})$\\
where $\chi_{\mu}$ and $\rho_{\mu\nu}$  have been restricted to $\Sigma$ and
are respectively even and odd
functions on the $r=$constant, $t=$constant 2-sphere for large $r$.

Finally, we may observe that all the arguments in the paper would go
through if we
imposed the opposite parity on $A_{a}$ from that expressed in (68).  Our
choice of sign was made only to conform with the character of the field due
to a static point charge at the origin of Minkowski spacetime.  It means in
effect that the spacetime inversion ${\cal I}$ has been chosen to act as
 $PT$ rather than as $CPT$.

\end{document}